\documentclass[preprint]{aastex}
\usepackage{graphics,graphicx}
\usepackage{subfigure}
\usepackage{epsfig}
\usepackage{appendix}
\usepackage{lscape}
\usepackage{longtable}
\usepackage{url}
\usepackage{bigstrut,bigdelim,multirow}

\shorttitle{Characteristics of Multi-episode Emission Patterns}
\shortauthors{Lan et al.}
\begin{document}
\title{Characteristics of Two-episode Emission Patterns in {\em Fermi} Long Gamma-Ray Bursts}
\author{Lin Lan\altaffilmark{1}, Hou-Jun L\"{u}$^{\ast}$\altaffilmark{1}, Shu-Qing Zhong\altaffilmark{1},
Hai-Ming Zhang\altaffilmark{1}, Jared Rice\altaffilmark{2}, Ji-Gui Cheng\altaffilmark{1}, Shen-Shi
Du\altaffilmark{1}, Long Li\altaffilmark{1}, Jie Lin\altaffilmark{1}, Rui-Jing Lu\altaffilmark{1}, and
En-Wei Liang\altaffilmark{1}} \altaffiltext{1}{Guangxi Key Laboratory for Relativistic Astrophysics,
Department of Physics, Guangxi University, Nanning 530004, China;
lhj@gxu.edu.cn}\altaffiltext{2}{Department of Physics and Astronomy, University of Nevada Las Vegas,
Las Vegas, NV 89154, USA}

\begin{abstract}
Two-episode emission components separated by quiescent gaps in the prompt emission of gamma-ray bursts
(GRBs) have been observed in the {\em Swift} era, but there is a lack of spectral information due to
the narrow energy band of the {\em Swift}/Burst Alert Telescope. In this paper, a systematic analysis
of the spectral and temporal properties of the prompt emission of 101 {\em Fermi}/Gamma-ray Burst
Monitor detected long GRBs show the existence of two-episode emission components in the light curves,
with quiescent times of up to hundreds of seconds. We focus on investigating the differences of those
two emission episodes. We find that the light curves of the two emission components exhibit different
behavior, e.g., a soft emission component that either precedes or follows the main prompt emission or
that the intensity of the two emission episodes are comparable with each other. No statistically
significant correlation in the duration of the two emission episodes can be claimed. We define a new
parameter $\varepsilon$ as the ratio of the peak flux of the first and second emission episodes and
find that a higher $\varepsilon$ corresponds to a larger fluence. The preferred spectral model in our
analysis is a cutoff power-law model for most GRBs. The distribution of $E_p$ for episodes I and II
range from tens of keV to 1000 keV with a lognormal fit and there are no significant differences
between them. Moreover, we do not find significant relationships between $\varepsilon$ and $E_p$ for
the two emission episodes. Those results suggest that these two-episode emission components likely
share the same physical origin.
\end{abstract}


\slugcomment{}
\section{Introduction}
Quantitatively, the duration of a burst is usually defined by $T_{90}$ which is the time interval
between 5\% and 95\% of the total fluence for a given detector. Phenomenologically, gamma-ray bursts
(GRBs) are classified into two categories, long and short, with a division line at the observed
duration $T_{90}\sim2$ seconds in the CGRO/BATSE era (Kouveliotou et al. 1993). The bimodal
distribution of $T_{90}$, on the other hand, together with information about host galaxies and
afterglow, suggests that long GRBs are related to the deaths of massive stars (Narayan et al. 1992;
Woosley 1993; MacFadyen \& Woosley 1999; Berger et al. 2005; Tanvir et al. 2005; Fruchter et al. 2006;
Zhang 2006), and short GRBs are associated with the merger of two compact stellar objects, such as the
merger of two neutron stars (Paczynski 1986; Eichler et al. 1989), or the merger of a neutron star and
a black hole (Paczynski 1991). These progenitors result in the immediate formation of a black hole (or
magnetar) that powers a relativistic jet, whose radiation can be observed if the jet is pointing toward
the direction of the observer from within an optically thin region (Usov 1992; Woosley 1993; Thompson
1994; Dai \& Lu 1998; Popham et al. 1999; Zhang \& M\'esz\'aros 2001; Metzger et al. 2008; Lei et al.
2013; L\"{u} \& Zhang 2014; Liu et al. 2017). The prompt emission and afterglow emission are usually
considered to be emission from internal shocks (ejecta internally dissipating energy) and external
shocks (ejecta interacting with a circum-burst medium), respectively (Paczynski 1986; Shemi \& Piran
1990; M\'esz\'aros \& Rees 1997; Kumar 1999; Kumar \& Zhang 2015).

Moreover, the duration of the prompt emission can reflect the activity of the GRB central engine (Zhang
et al. 2016). However, Qin et al. (2013) found that the bimodal $T_{90}$ distribution of {\em Fermi}/
Gamma-ray Burst Monitor (GBM) data is significantly different from the distributions found in both {\em
BeppoSAX}/GRBM and {\em Swift}/Burst Alert Telescope (BAT) data. They also found differences in the
measured $T_{90}$ in different energy bands for the same burst. These findings indicate that $T_{90}$
is significantly affected by instrumental selection effects and is energy-dependent. On the other hand,
the long, soft extended emission tails (Della Valle et al. 2006; Fynbo et al. 2006; Gal-Yam et al.
2006; Gehrels et al. 2006), the internal plateau (Troja et al. 2007; Lyons et al. 2010; Rowlinson et
al. 2010, 2013; L\"{u} et al. 2015), and the erratic, bright X-ray flares observed with {\em Swift}
/BAT and the X-ray Telescope (XRT) challenge the simple classification based on the $T_{90}$ criterion
(Barthelmy et al. 2005; Nousek et al. 2006; Zhang et al. 2006; Perley et al. 2009; L\"{u} et al. 2010;
Zhang et al. 2014; Gao et al. 2017). In particular, the X-ray flares and internal plateaus observed in
XRT are very difficult to interpret with the external shock model. The late activities of the GRB
central engine (Burrows et al. 2005; Fan \& Wei 2005; King et al. 2005; Dai et al. 2006; Zhang et al.
2006) show that internal dissipation is more favorable in explaining these phenomena. This
observational evidence suggests that the activity of the central engine of GRBs does not stop for
longer than $10^4$ s.

Observational data indicate that at least $\sim(9-15)\%$ of GRB prompt emission signatures are composed
of two or more emission episodes with a quiescent time that may be extremely long, e.g., up to
$\sim$100 s in the rest frame (Koshut et al. 1995; Lazzati 2005; Burlon et al. 2008; Bernardini et al.
2013; Hu et al. 2014). Considering the case of two-episode emission components, a single prompt
emission light-curve pattern resembling a soft emission (precursor)does not exist prior to the main
prompt event (Koshut et al. 1995; Lazzati 2005; Burlon et al. 2008; Hu et al. 2014), nor does a soft
emission (similar to extended emission) after the main prompt event, or the emissions from those two
episodes are comparable with each other. Beyond that, some events observed by {\em Swift}/BAT were
triggered twice with a temporal separation of hundreds of seconds, called double bursts (e.g., GRB
110709B; Zhang et al.2012; Hu et al. 2014; Liu et al. 2014). On the other hand, in previous studies,
the spectral analysis of those two-episode emissions used the hardness ratio to roughly compare the
spectral properties. However, this cannot reflect the intrinsic properties of the spectra due to the
narrow energy band of {\em Swift}/BAT (Barthelmy et al. 2005). Therefore, measuring the peak energy
($E_{p}$) of spectra is an important aspect of understanding the hardness of the two-episode emissions
and determining if there are intrinsic relationships and evolution between them.

It is possible to diagnose the spectral properties of those multiepisode emissions thanks to the {\em
Fermi} satellite, which provides an unprecedented spectral coverage over 7 orders of magnitude in
energy (from $\sim$8 keV to $\sim$300 GeV). {\em Fermi} includes two instruments. One is the GBM
containing 12 sodium iodide (NaI) detectors that cover an energy range from 8 keV to 1 MeV and two
bismuth germanate (BGO) scintillation detectors that are sensitive to higher energies between 200 keV
and 40 MeV (Meegan et al. 2009). The other is the Large Area Telescope (LAT) with an energy coverage
from 20 MeV to 300 GeV (Atwood et al. 2009). Observationally, not all GRBs detected by GBM are seen
with high-energy emission in LAT and only a few bursts have exhibited high-energy emission over the
last 9 years (Ackermann et al.2013; Vianello et al. 2015). Due to the narrow energy band of {\em Swift}
/BAT, it is often not possible to derive a full spectrum, and the best fit is a simple power-law (PL)
model. Nevertheless, {\em Fermi}/GBM covers a broad energy band, and the peak energy ($E_{p}$) of the
spectrum of most GRB prompt emission episodes can be derived by invoking a Band function, or cutoff
power-law (CPL) model. Several questions regarding these multiepisode emission of prompt emission GRBs
are worthy of discussion. Does a spectral evolution exist between the emission episodes? Are there
correlations between observed parameters for the episodes? Our goal is to systematically analyze the
GRB data observed by {\em Fermi}/GBM since its operations began in 2008, aiming to address the
abovementioned problems in GRB prompt emission. In this paper, we focus on a comprehensive analysis of
the GRBs detected by GBM, and present a two-episode emission GRB catalog with our spectral and temporal
analysis. Our data reduction and analysis methods are presented in \S2 and the analysis results are
reported in \S3. The conclusions and a discussion are presented in \S4.

\section{Data reduction and sample selection}
\subsection{Light-curve extraction}
{\em Fermi}, a high-energy $\gamma$-ray satellite, has operated for more than 9 years since it was
launched in 2008. There are three different types of signals from each of the 14 GBM detectors: CTIME,
CSPEC, and TTE\footnote{The continuous time (CTIME) data include eight energy channels and have a finer
time resolution of 64 ms. The continuous spectroscopy (CSPEC) data include 128 energy channels, and a
time resolution of 1.024 s. The time-tagged event (TTE) data consist of individual detector events,
each tagged with arrival time, energy (128 channels), and detector number (Paciesas et al. 2012).}. We
do not use CTIME data and CSPEC data in our analysis due to the fixed time resolution of 64 ms and
1.024 s, but select the TTE data that include individual photons arriving with time and energy tags.
Moreover, we can select any time resolution bin size to perform our spectral and temporal analysis.

We download the original GBM data (12 NaI and 2 BGO detectors), as well as LAT data of GRBs from the
public science support center at the official {\em Fermi}
website\footnote{http://fermi.gsfc.nasa.gov/ssc/data/.}. We choose the brightest detector among NaI and
BGO to do the analysis, respectively, because the brightest detector has a minimum angle between the
incident photon and the normal direction of the detector. Based on the standard \texttt{heasoft} tools
(version 6.19) and the {\em Fermi} \texttt{ScienceTools} (v10r0p5), a {\em Python} code was developed
to extract the energy-dependent light curves and time-dependent spectra using the spectral source
package $gtBurst$\footnote{http://sourceforge.net/projects/gtburst/.}. We employ the Bayesian Block
algorithm to identify the light curves. Please refer to our latest paper ( Lu et al. 2017) for more
details on data analysis with the Bayesian Block algorithm. Several points need to be cautioned:
firstly, we extract the light curve with a time bin of 128 ms to identify a possible signal in
different energy bands in the time interval [-100, 300] s, and we adopt 10 control blocks for picking
up a weaker signal in our analysis. Then, we extract the light curve again by adopting a 64 ms time bin
instead of the 128 ms time bin, if a possible Bayesian block can be identified. The motivation for
using a 64 ms time bin by running again instead of a 128 ms time bin is to find microstructure in the
light curves, in particular, in the soft precursor emission, which may have a short duration. Also, a
signal-to-noise ratio (S/N) of at least 3$\sigma$ is believed to be a true signal \footnote{In order to
obtain the arrival time of different energy photons, we separate the NaI and BGO detectors into two
energy bands, respectively, e.g., [8, 50] keV and [50, 1000] keV for NaI; [250, 1000] keV and $>$1000
keV for BGO.}. Finally, we calculate the duration (5\% and 95\% of the total photons for each episode)
of the bursts within the energy band 8 - 1000 keV. In Table 1, we summarize our results and indicate
$T_{\rm E,I}$ and $T_{\rm E,II}$ as the duration of the first and second emission episodes.

\subsection{Sample selection criteria}
As of August 2017, we have extracted the light curves of 2059 GRBs that were detected by {\em Fermi}
/GBM. The LAT light curve is also extracted if the LAT data are available. The full data set includes
1764 long GRBs and 295 short GRBs. There are three criteria adopted for our sample selection. First,
the prompt emission must have two emission episodes with the quiescent time between these
episodes being longer than 5 s. Second, the S/N ratio of the emission episodes should be greater than
3$\sigma$. Third, we focus on long GRBs that have a main prompt emission duration longer than 5 s.
There are 101 GRBs that satisfy our criteria up to 2017 August, 4 of which have a measured redshift,
e.g., GRBs 091208B, 100615A, 140512A, and 151027A. It also includes 6 GRBs detected by LAT, e.g. GRBs
090328, 100116A, 121225B, 130821A, 150118B, and 150523A.

\subsection{Spectrum extraction}
\begin{figure}
\begin{tabular}{lllll}
\multirow{2}{*}{ \includegraphics[angle=0,scale=0.40]{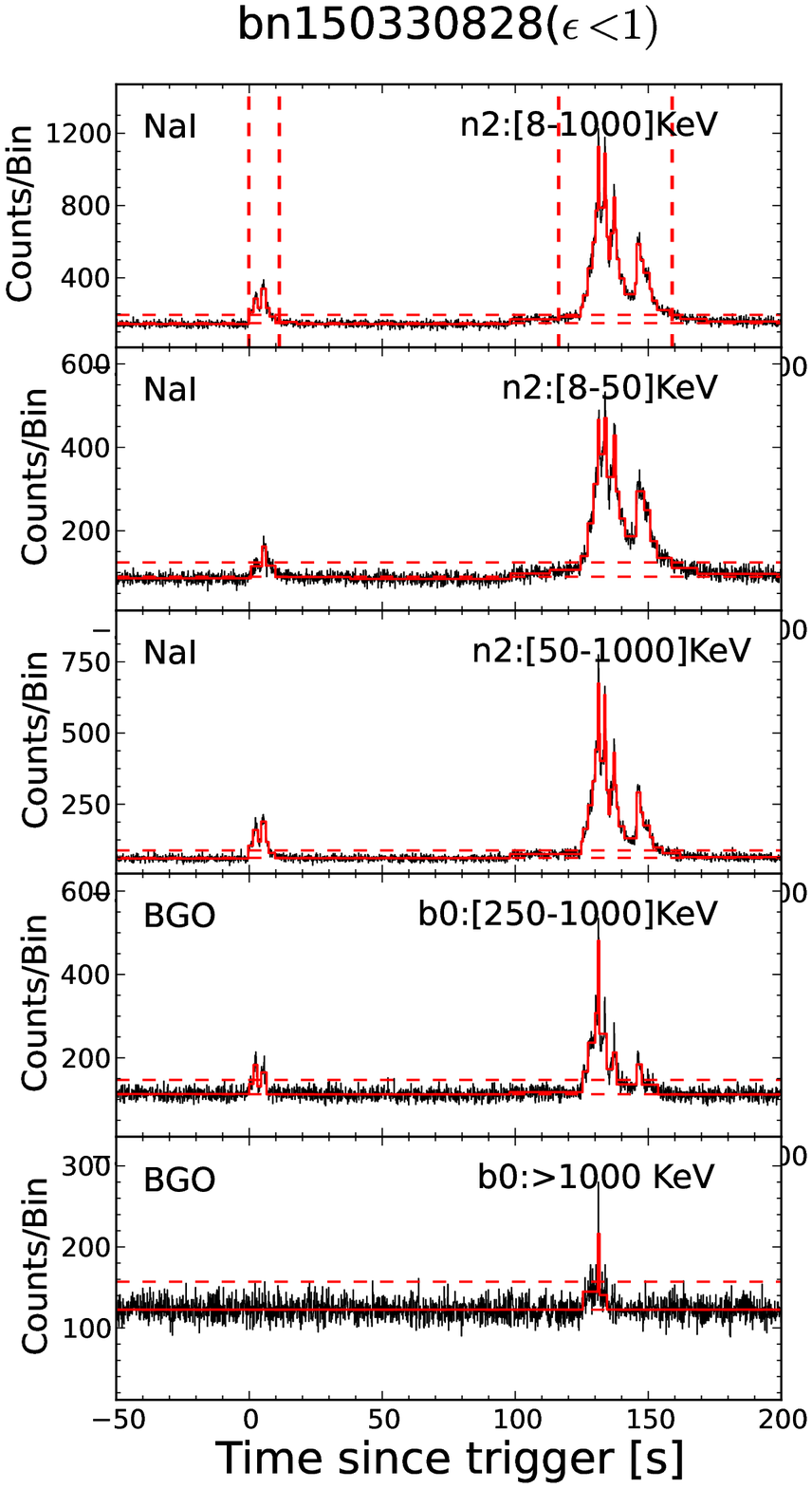}} \\
&\includegraphics[angle=270,scale=0.22]{bn150330828_1.eps} \\
&\includegraphics[angle=270,scale=0.22]{bn150330828_2.eps} \\
\multirow{2}{*}{ \includegraphics[angle=0,scale=0.40]{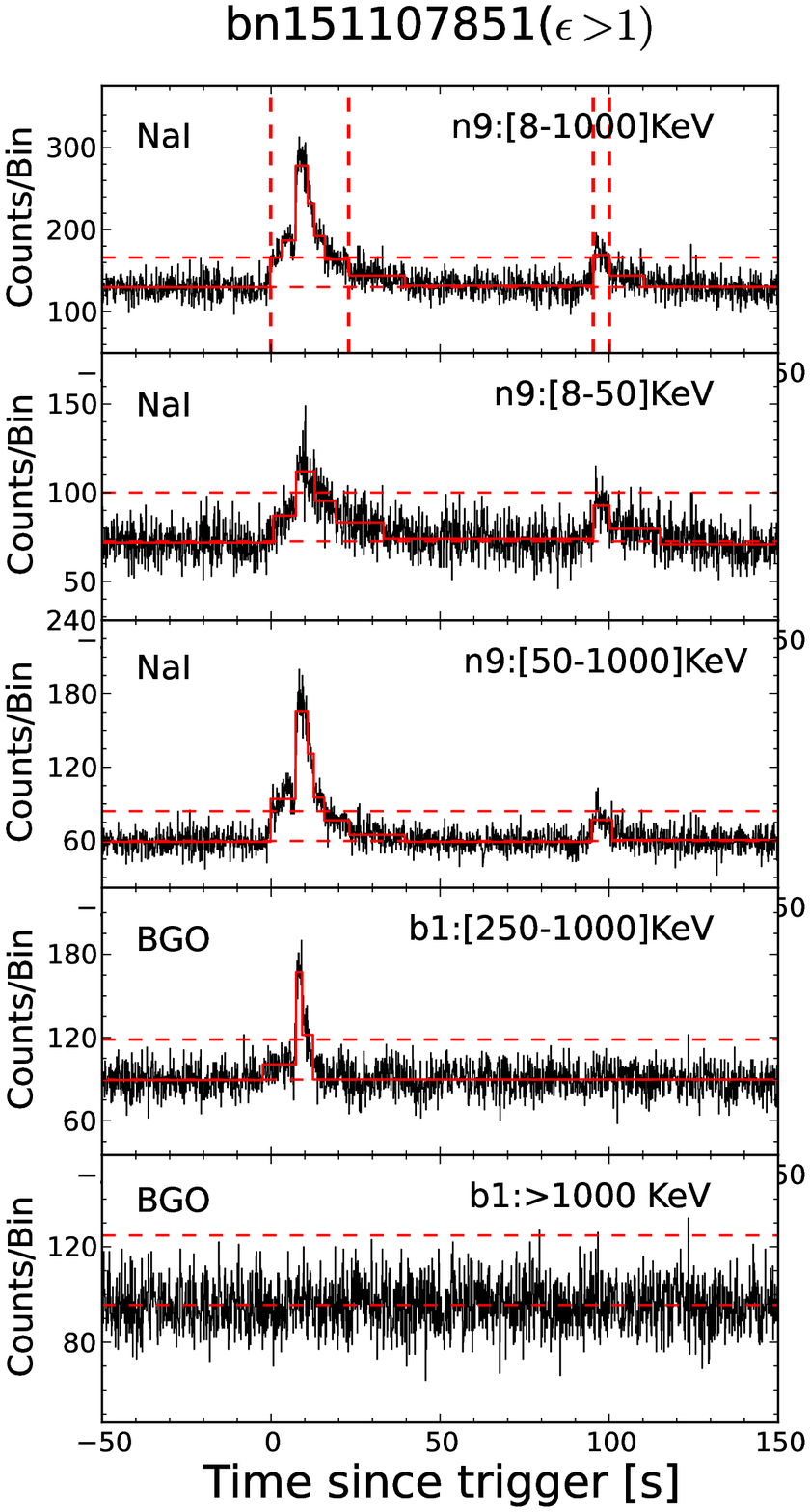}} \\
&\includegraphics[angle=270,scale=0.22]{bn151107851_1.eps} \\
&\includegraphics[angle=270,scale=0.22]{bn151107851_2.eps} \\
\end{tabular}
\caption{Three examples of two episodes' emission light curves (e.g., episode I is brighter than episode
II, episode II is brighter than episode I, and almost equal episodes) and spectra, together with our
Bayesian block analysis (red blocks in the left panels) and spectral fits for each episode
(solid line in the right panels). The dashed horizontal lines in the left panels are 3$\sigma$ signal
over background emission. The dashed vertical lines are the beginning and end of each emission episode.}
\label{LC2example}
\end{figure}

The background spectrum from the GBM data are extracted from two time intervals: before the first
emission episode and after the last emission episode, and are modeled with a polynomial function. XSPEC
is used to perform spectral fits for each episode. The motivation of this work is focused on comparing
the difference between those two episodes' emission. So that, in our spectral analysis, we only perform
the time-integrated spectral fits even though the spectral evolution may exist for each episode
emission. The statistic $\chi^2$ is adopted to judge the goodness of the spectral fits. In our analysis
a Band function model (Band et al. 1993), a widely used phenomenological model, is invoked as the
primary model and is written as
\begin{eqnarray}
N_{\rm Band}(E)=B(E)=A\cases{(\frac{E}{100 \rm~keV})^{\alpha}\rm exp(-\frac{E}{E_0}),
            & $E<(\alpha-\beta)E_0$ \cr
[\frac{(\alpha-\beta)E_0}{100 \rm~keV}]^{\alpha-\beta} \rm exp(\beta-\alpha)(\frac{E}{100 \rm~keV})^{\beta},
            & $E\geq (\alpha-\beta)E_0$ \cr            }
\end{eqnarray}
where $A$ is the normalization of the spectrum, $\alpha$ and $\beta$ are the low and high-energy photon
spectral indices, respectively; $E_0$ is the break energy of the spectrum, and the peak energy
($E_{p}$) of the spectrum is related to $E_0$ through
\begin{eqnarray}
 E_p = (2+\alpha)E_{0}.
\end{eqnarray}
Alternatively, if the Band function model is not good enough to fit the data, a CPL or a simple PL
model are adopted, which can be written as
\begin{eqnarray}
N_{\rm CPL}(E) = A\cdot E^{-\alpha}\rm exp(-\frac{E}{E_{\rm p}}),
\end{eqnarray}
\begin{eqnarray}
N_{\rm PL}(E) = A\cdot E^{-\alpha}.
\end{eqnarray}
The examples of spectral fitting in our sample are shown in Figure \ref{LC2example}, and the fitting
results for each episode are shown in Table 1, The peak energy and PL indices of the spectrum of the
first and second emission episodes are indicated as  $E_{p, \rm I}$ and $E_{p, \rm II}$, and as
$\alpha_{\rm I}$ and $\alpha_{\rm II}$, respectively.

\section{Results}
\subsection{Light curves, duration, and quiescent time}
Phenomenologically, the light curves of those two-episode emission components show different behaviors,
e.g., a soft emission component prior to (40 out of 101) or following (7 out of 101) the main prompt
emission,\footnote{Here, we define the peak flux of each episode to indicate the strength of each
emission episode.} or the intensity of the two emission episodes are comparable with each other (54 out
of 101). Three examples of such light-curve patterns are shown in Figure \ref{LC2example}.

\begin{figure}
\centering
\includegraphics[angle=0,width=0.8\textwidth]{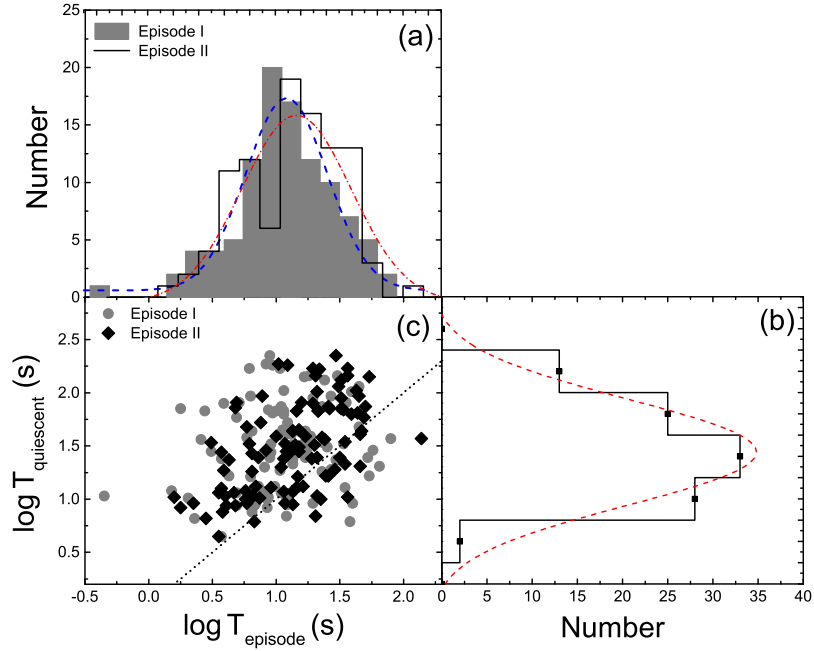}
\caption{1D and 2D distributions for the two-episode emissions in our sample in the $\log T_{\rm
quiescent}$ vs. $\log T_{\rm episode}$. The duration distributions of the first (I) and second (II)
episodes' emission are shown in panel (a), while the duration of the quiescent time is shown in panel (b).
The blue dashed line, red dashed-dotted and dashed lines are the best Gaussian fit. Panel (c) displays $T_{\rm
quiescent}$ as a function of the duration of episode I (gray dots) and II (black diamonds). The black
dotted line corresponds to $T_{\rm quiescent}=T_{\rm episode}$.}
\label{Duration}
\end{figure}

Based on the Bayesian block method, we calculate the duration of each emission episode, ($T_{\rm
episode}$), and show the distribution of $T_{\rm episode}$ in Figure \ref{Duration}(a), as well as the
distribution of quiescent time\footnote{The definition of quiescent time is the duration between the
end of the first episode and the beginning of the second episode.}($T_{\rm quiescent}$) between the two
emission episodes in Figure \ref{Duration}(b). The duration of the two emission episodes and the
quiescent time follow a lognormal distribution with $\log T_{\rm episode, I}=(1.09\pm 0.02)$ s, $\log
T_{\rm episode, II}=(1.16\pm 0.05)$ s, and $\log T_{\rm quiescent}=(1.44\pm 0.05)$ s, respectively. The
maximum and minimum quiescent times in our sample are 223.36 s for GRB 131108A and 5.42 s for GRB
091208B, respectively. Bernardini et al. (2013) proposed that the precursor and prompt emission arise
from the accretion of matter onto the surface of the magnetar, and a longer quiescent time of the two
emission episodes should correspond to a higher intensity for the second emission episode. However, we
do not find any correlations between $T_{\rm quiescent}$ and $T_{\rm episode, I/II}$. This suggests
that the physical origin of the two emission episodes may not be related to the accretion physics.
Moreover, it may introduce constraints on the proposed models that are invoked for interpreting the two
emission episodes.

\begin{figure}
\centering
\includegraphics[angle=0,width=0.8\textwidth]{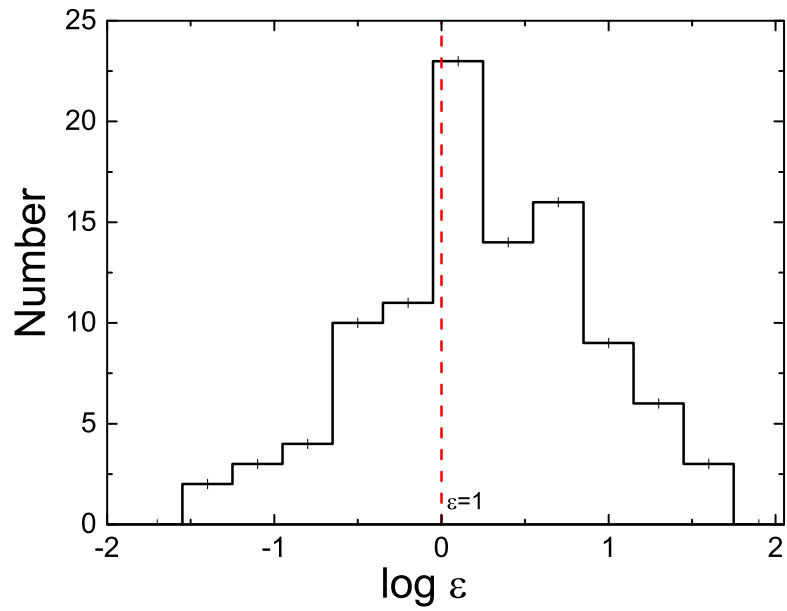}
\caption{Distribution of $\log \varepsilon$ for two-episode emissions in our sample. The
definition of $\varepsilon$ is the ration of the 1 s peak flux between the first and
second episodes' emission within 8 keV - 40 MeV. The dashed vertical line is $\varepsilon=1$.}
\label{Parameter}
\end{figure}

\begin{figure}
\centering
\includegraphics[angle=0,width=0.8\textwidth]{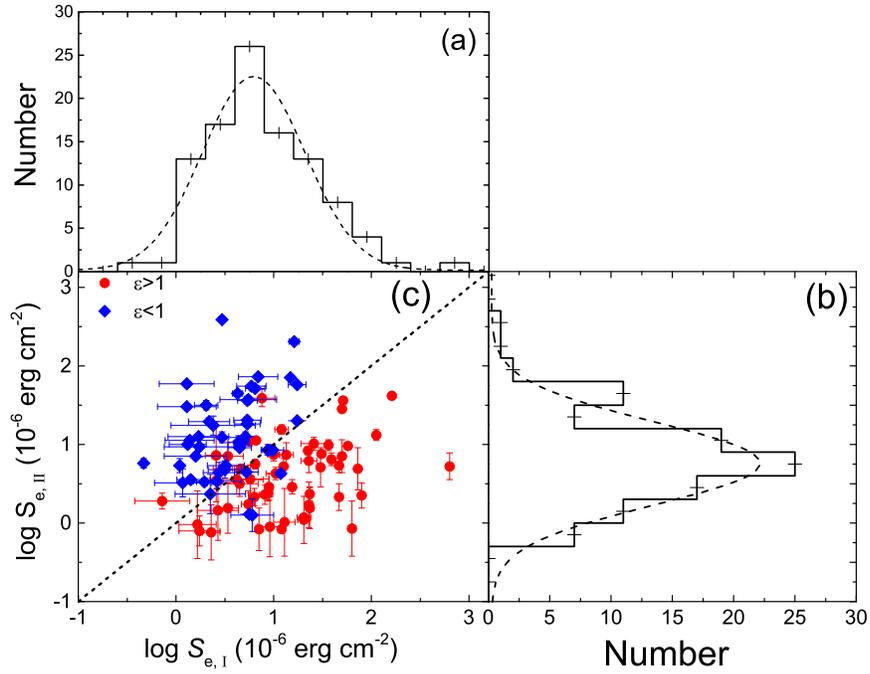}
\caption{Fluence distributions of the first (I) and second (II) episodes'
emission are shown in panels (a) and (b), respectively. The dashed lines are the best Gaussian fits.
Panel (c) shows the correlation of fluence for the first (I) and second (II) episodes'
emission. The dotted line is $S_{\rm e, I}=S_{\rm e, II}$. The blue diamonds
and red points correspond to $\varepsilon<1$ and $\varepsilon>1$, respectively.}
\label{Fluence}
\end{figure}

In order to identify the brightness of each episode from the same burst, we define one dimensionless
parameter, the relative factor $\varepsilon$, which is the ratio of the 1-s peak flux between the two
emission episodes within 8 keV - 40 MeV,
\begin{eqnarray}
\varepsilon = \frac{F_{\rm e,I}}{F_{\rm e,II}},
\end{eqnarray}
where $F_{\rm e,I}$ and $F_{\rm e,II}$ are the 1 s peak flux of the first and second episodes,
respectively. We extract the 1 s peak flux of each episode based on the spectra and present them in
Table 1. Figure \ref{Parameter} shows the distribution of $\varepsilon$ in our sample. Figures \ref
{Fluence}(a) and (b) present the distributions of fluence for episodes I and II in our sample, which
follow a lognormal distribution with $\log S_{\rm e, I}=(-5.21\pm 0.04) \rm~erg~cm^{-2}$ and $\log
S_{\rm e, II}=(-5.24\pm0.05) \rm~erg~cm^{-2}$. Interestingly, we found that larger $\varepsilon$ (the
first emission episode brighter) corresponds to a higher fluence, see Figure \ref{Fluence}(c).

\subsection{Spectral properties}
Although the Band function (Band) is our primary spectral model for fitting the data and the $\chi^2$
statistic is good enough, the high-energy index $\beta$ cannot be constrained very well. The values of
$\beta$ have large error bars and are much less than the typical value $-2.3$ for most GRBs. There are
no significant differences in the fits with a Band model as compared with {the CPL model, but the CPL
model has fewer independent parameters. The $\chi^2$ value of the PL model fits} is much larger than
that of the CPL models. Beyond that, we also try to invoke the blackbody (BB) model or more free
parameter multi-component models to fit the spectra (e.g., BB+Band, BB+CPL, and BB+PL), but the
temperature of BB is very difficult to constrain. Therefore, the preferred spectral model in our
analysis is the CPL model for most GRBs, except for two episodes that are modeled well by a Band
function, e.g. the second episode of GRB 150330A (bn150330828) and the first episode of GRB 170409A
(bn170409112). On the other hand, LAT data are also presented for six GRBs in our sample. By
considering the contributions of LAT data for spectral analysis in the two emission episodes, the first
and second episodes of GRBs 121225B, 130821A, and 150118B are well modeled by a Band function and a CPL
model, respectively. Moreover, the first emission episodes of GRBs 090328 and 150523 are well modeled
by a Band function with an extra CPL or PL components, as well as the second episode emission of GRB
100116A with the Band function model. The spectral parameters derived from our fits are reported in
Table 1.

\begin{figure}
\centering
\includegraphics[angle=0,width=0.8\textwidth]{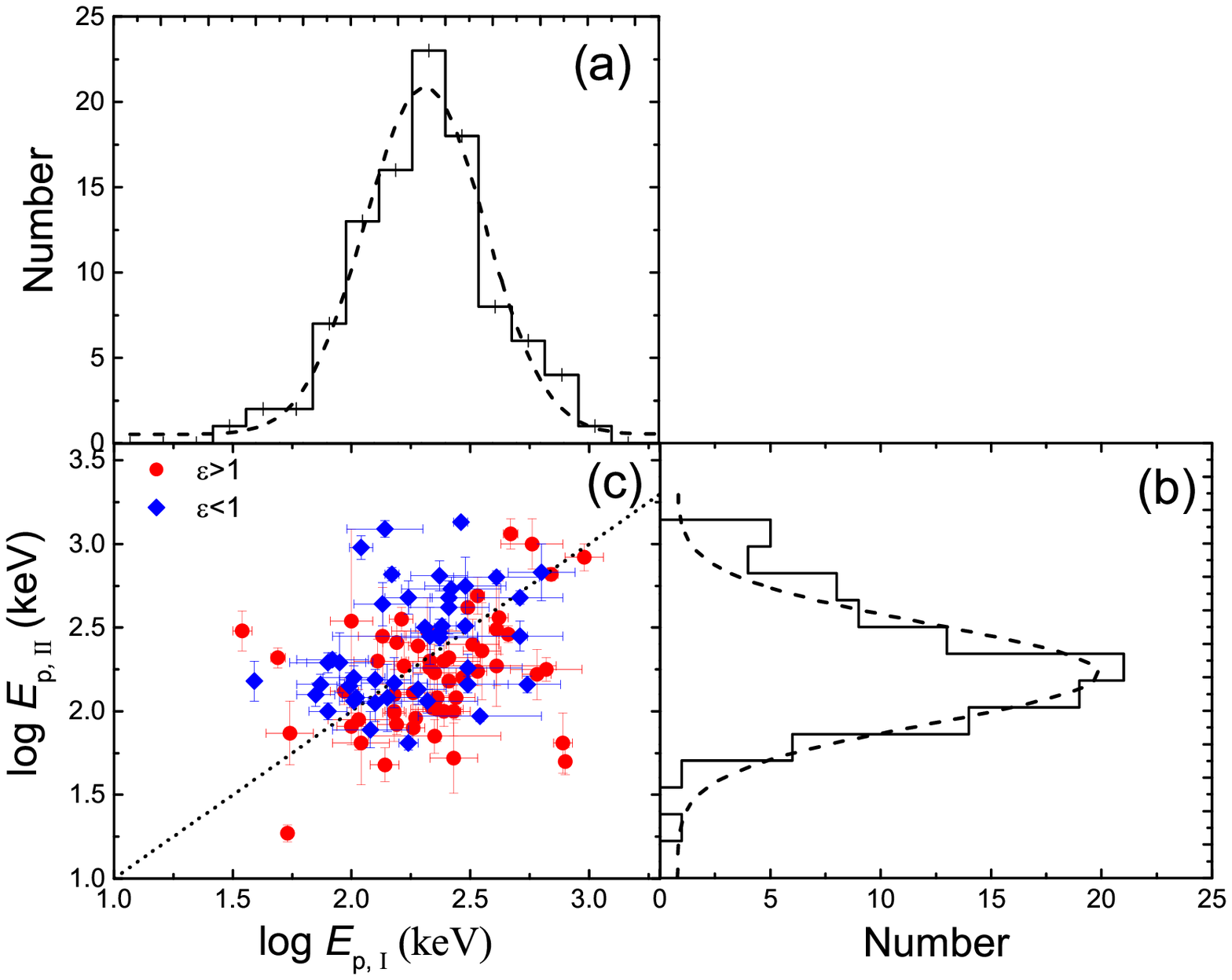}
\includegraphics[angle=0,width=0.8\textwidth]{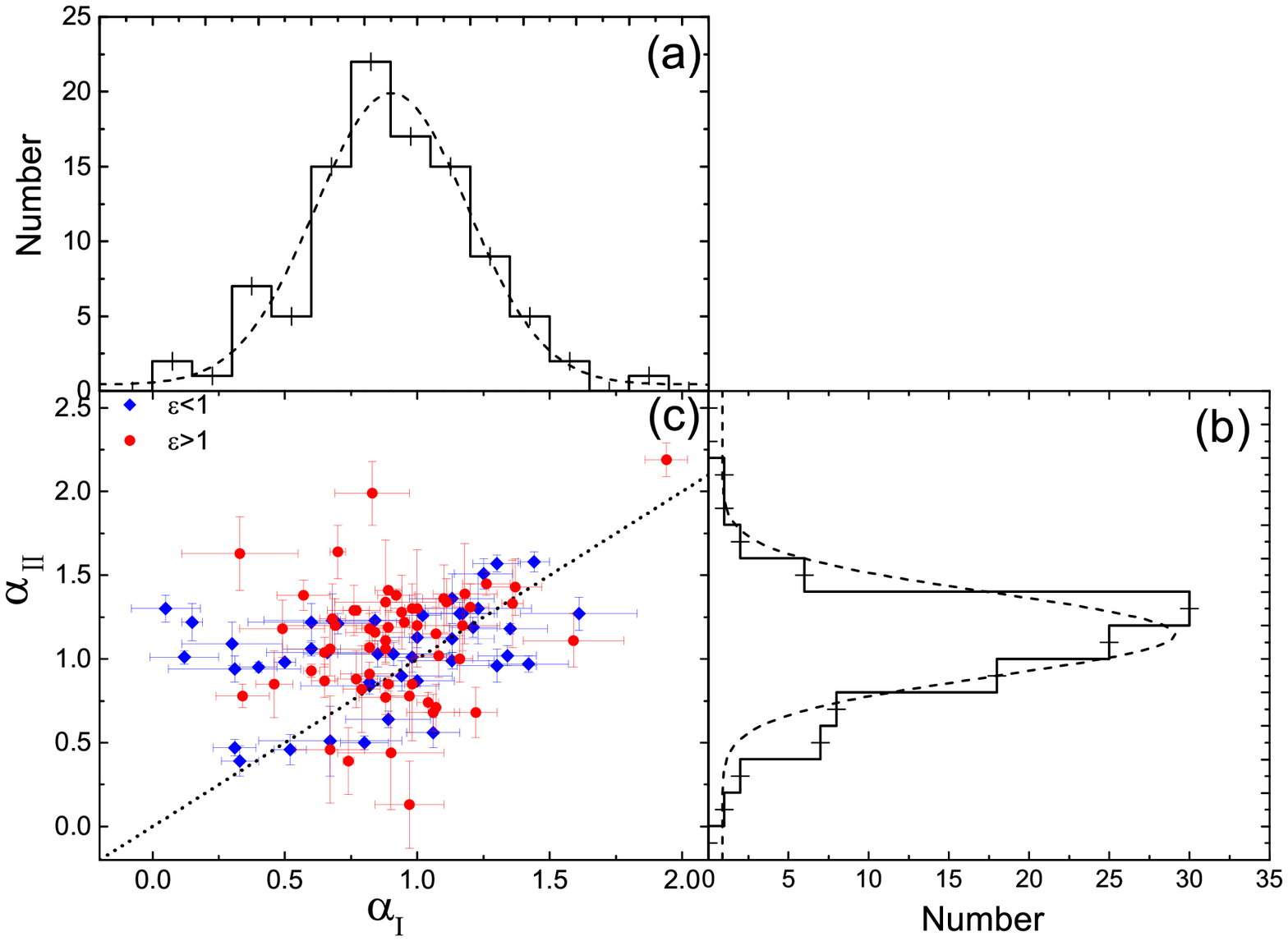}
\caption{Distributions of peak energy $E_{p, I}$ and $E_{p, II}$ (top panel),
photon index $\alpha_{\rm I}$ and $\alpha_{\rm II}$ (bottom panel), and their correlations for our
sample. Black dashed lines are the best Gaussian fits, and black dotted lines correspond to
$E_{\rm p, I}=E_{\rm p, II}$ (top panel) and $\alpha_{\rm I}=\alpha_{\rm II}$ (bottom panel).
The red dots and blue diamonds correspond to $\varepsilon>1$ and $\varepsilon<1$, respectively.}
\label{Spectra}
\end{figure}

However, the origin of high-energy GeV photons remains under debate, as well as whether keV - MeV and
GeV photons share the same physical origin. Thus, by ignoring the contribution of LAT/LLE data, we
reanalyze the spectra of those six GRBs with LAT detections, and find that the CPL model is favored to
perform spectral fits. The top panels of Figure \ref{Spectra} ((a) and (b)) show the distributions of
$E_{p, \rm I}$ and $E_{p, \rm II}$, the peak energy of the spectra for the first and second emission
episodes. The $E_{p, \rm I}$ and $E_{p, \rm II}$ distributions range from tens of keV to 1000 keV. Both
of them are followed by lognormal distributions with peaks at $\log E_{\rm p, I}=(2.31\pm0.02)$ keV and
$\log E_{\rm p, II} =(2.22\pm0.03)$ keV, respectively. Similarly, the bottom panels of Figure
\ref{Spectra} ((a) and (b)) show the distributions of $\alpha_{\rm I}$ and $\alpha_{\rm II}$, the
spectral index for the first and second episodes' emission. The distributions are also normal with mean
values of $\alpha_{\rm I}=0.9\pm 0.02$ and $\alpha_{\rm II}=1.15\pm0.03$. From a statistical point of
view, the peak energy of the first emission episode appear to be slightly harder than that of the
second episode. However, a statistical test using the method proposed by Ashman et al. (1994) yields
$P_{\rm ks} = 0.083$, which indicates that they cannot be absolutely distinguished as arising from a
different population.

In order to test whether the brighter emission episode corresponds to a higher $E_p$ of the spectrum
forthe same burst, Figure \ref{Spectra} (c) shows the distributions of $E_{p, \rm I}-E_{p, \rm II}$ and
$\alpha_{\rm I}-\alpha_{\rm II}$. We do not find a significant difference between higher (or lower)
$\varepsilon$ and harder (or softer) $E_p$. A similar result holds for the photon index $\alpha$ of
spectral fits.

\section{Conclusions and Discussion}
We have presented a comprehensive temporal and spectral analysis for two-episode prompt emission of
GRB data observed with {\em Fermi}/GBM during 9 years of operation, where the two emission episodes are
separated by a quiescent time of up to hundreds of seconds. We studied the possibility of the existence
of significant differences between the two emission episodes. Our results are summarized as follows:
\begin{itemize}
\item We find that the light curves of 101 long GRBs with two-episode emission components show
    different behavior, e.g., a soft emission component prior to (40 out of 101) or following (7
    out of 101) the main prompt emission, or that the intensity of those two emission episodes are
    comparable (54 out of 101).
\item The distributions of episodes I, II, and the quiescent time are normal. The maximum and
    minimum quiescent times in our sample are 223.36 s for GRB 131108A and 5.42 s for GRB 091208B,
    respectively. There are no correlations between $T_{\rm quiescent}$ and $T_{\rm episode,
    I/II}$, but there is a possible correlation between $\varepsilon$ and fluence, e.g., the
    brighter episode corresponds to higher fluence.
\item The preferred spectral model in our analysis is the CPL model for most GRBs. The distribution
    of $E_p$ for episodes I and II are lognormal and range from tens of keV to 1000 keV. Moreover,
    we do not find significant differences between higher (or lower) $\varepsilon$ and harder (or
    softer) $E_p$. This suggests that the two-episode emission components may share the same
    physical origin.
\end{itemize}

Based on our analysis, about 5\% of long GRBs include two-episode emission components for the same
burst in the {\em Fermi} era (101 out of 1764). This fraction is less than that in the {\em Swift} era
($\sim$9\%, Koshut et al. 1995; Lazzati 2005; Burlon et al. 2008; Bernardini et al. 2013; Hu et al.
2014). This difference may be related to the different method we used to search for the signal or the
fact that the sensitivity of different instruments with different trigger energy maybe exibit
differences. On the other hand, there are two ways by which such two-episode emission may be caused in
two ways: either the emission is continuous but the signal in the quiescent time is too weak to be
detected by current $\gamma$-ray detectors, e.g. the tip-of-the-iceberg effect in L\"{u} et al. 2014 or
the emission is intrinsic and no signal exists during the quiescent time (Hu et al. 2014).

Lu et al. (2012) showed either an evolution from hard to soft for single-pulse or multipulse GRBs
detected by {\em Fermi}/GBM or intensity tracking behavior. In order to test whether the soft first
emission episode is the subemission of the second one, we compare the peak energy of the first emission
episode with the second. In our analysis, we do not find a difference in the $E_p$ distribution between
the two episodes, indicating that the first episode is not likely to be subemission of the second
episode.

From a theoretical point of view, the physical interpretation of these two-episode emissions remains an
open question. One possible interpretation is that when a jet propagates within the stellar envelope
(Ramirez-Ruiz et al. 2002; Lazzati \& Begelman 2010; Nakar \& Piran 2017) the wasted energy of the jet
is recycled into a high pressure cocoon surrounding the relativistic jet (Ramirez-Ruiz et al. 2002;
Lazzati \& Begelman 2005). Wang \& M\'esz\'aros (2007) proposed that the first soft emission is
produced by the jet bow shock, the quiescent time is due to the pressure drop ahead of the jet head
after it reaches the stellar surface, and the second emission episode is from a relativistic jet that
is accelerated by a rarefaction wave. Lipunova et al. (2009) proposed two steps in the collapse of the
progenitor star, i.e., it first collapses to a neutron star (soft emission) and then into a black hole
(main emission). These models predict that the time gap is only tens of seconds and the first soft
emission has a thermal spectrum. Even a small fraction of the time gap for our samples ranges over tens
of seconds, but the lack of a thermal component for spectra from observations is not consistent with
the prediction of the above models. Alternatively, the multiepisode nature of these events may be
directly related to central engine activity (Ramirez-Ruiz et al. 2001) similar to a newly born magnetar
as a the central engine of GRBs, the precursor and the prompt emission arise from the accretion of
matter onto the surface of the magnetar (Bernardini et al. 2013). However, it is difficult to interpret
the case of main emission followed by a soft emission. For the case of soft emission at a later time,
several models have been proposed, e.g., the collapse of a rapidly rotating stellar core leading to
fragmentation (King et al. 2005), fragmentation in the accretion disk (Perna et al. 2006), or a
magnetic barrier around the accretor (Proga \& Zhang 2006).

On the other hand, some possible interpretations of the double bursts are that a single GRB is located
behind a foreground galaxy and is gravitationally lensed before reaching the detectors, or the jet
precesses in a black hole hyper-accretion system (Zhang et al. 2012; Liu et al. 2014). These
interpretations require more observational data and information about the host galaxy to be identified.
Moreover, there are only four GRBs, GRBs 091208B, 100615A, 140512A, and 151027A, with measured
redshifts in our sample. Due to a lack of host galaxy information and optical observations in our
sample (leading to a lack of redshift measurements), one cannot measure an intrinsic peak luminosity
and isotropic energy for the two emission episodes in order to constrain the theoretical model. More
observational data are needed in the future to present a unified picture and interpret prompt emission
light curves.

\begin{acknowledgements}
We acknowledge the use of the public data from the {\em Fermi} data archive. This work is supported by
the National Basic Research Program (973 Programme) of China 2014CB845800, the National Natural Science
Foundation of China (grant Nos. 11603006, 11851304, 11533003, 11363002, and U1731239), the Guangxi
Science Foundation (grant Nos. 2017GXNSFFA198008, 2016GXNSFCB380005, and AD17129006). The
One-Hundred-Talents Program of Guangxi colleges, the high level innovation team and outstanding scholar
program in Guangxi colleges, Scientific Research Foundation of Guangxi University (grant No.
XGZ150299), and special funding for Guangxi distinguished professors (Bagui Yingcai and Bagui Xuezhe).

\end{acknowledgements}

\clearpage
\begin{deluxetable}{cccccccccccccc}
\centering \tabletypesize{\scriptsize} \rotate \tablecaption{{\bf Results of the temporal and spectral
analysis of 101 GRBs in our sample.}} \tablewidth{0pt} \tablehead{ \colhead{Trigger ID}&
\colhead{$T_{E,\uppercase\expandafter{\romannumeral1}}^{a}$}&
\colhead{$F_{p,\uppercase\expandafter{\romannumeral1}}^{b}$}&
\colhead{$E_{p,\uppercase\expandafter{\romannumeral1}}^{c}$}&
\colhead{$\alpha_{\uppercase\expandafter{\romannumeral1}}^{d}$} &
\colhead{$(\chi^{2}/$dof)}&
\colhead{$T_{q}^{e}$}&
\colhead{$T_{E,\uppercase\expandafter{\romannumeral2}}^{a}$}&
\colhead{$F_{p,\uppercase\expandafter{\romannumeral2}}^{b}$}&
\colhead{$E_{p,\uppercase\expandafter{\romannumeral2}}^{c}$}&
\colhead{$\alpha_{\uppercase\expandafter{\romannumeral2}}^{d}$}&
\colhead{$(\chi^{2}/$dof)}\\
\colhead{}&\colhead{(s)}&\colhead{($10^{-7} \rm
{erg~cm^{-2}~s^{-1}}$)}&\colhead{(keV)}&\colhead{}&\colhead{}&\colhead{(s)}&\colhead{(s)}&\colhead{($10^{-7}
\rm {erg~cm^{-2}~s^{-1}}$)}&\colhead{(keV)}&\colhead{}&\colhead{}}

\startdata

bn080724401	&	8.77 	&	6.44$\pm$3.94	&	103.25$\pm$13.76	&	0.80$\pm$0.09	&	257/237	&	10.54 	
&	1.60 	&	7.81$\pm$5.64	&	114.58$\pm$5.35	&	
0.50$\pm$0.04	&	241/237	\\
bn081009140	&	9.15 	&	73.02$\pm$2.15	&	53.18$\pm$1.75	&	1.11$\pm$0.03	&	260/233	&	27.86 	&	
12.80 	&	9.96$\pm$5.71	&	18.8$\pm$2.07	&	
1.34$\pm$0.14	&	256/238	\\
bn090131090	&	9.66 	&	21.66$\pm$2.03	&	48.79$\pm$3.75	&	0.92$\pm$0.07	&	267/236	&	13.23 	&	
16.26 	&	12.39$\pm$2.45	&	209.98$\pm$29.77	&	
1.38$\pm$0.04	&	242/237	\\
bn090328401	&	29.82 	&	14.18$\pm$5.80	&	767.59$\pm$78.09	&	0.98$\pm$0.03	&	253/237	&	23.23 	
&	4.29 	&	4.16$\pm$0.09	&	64.84$\pm$26.99	&	
0.85$\pm$0.34	&	247/237	\\
bn090524346	&	16.32 	&	12.98$\pm$5.52	&	99.43$\pm$12.82	&	0.76$\pm$0.10	&	220/239	&	27.01 	&	
6.53 	&	3.18$\pm$0.48	&	81.34$\pm$20.95	&	
1.29$\pm$0.15	&	249/239	\\
bn090529564	&	2.29 	&	17.68$\pm$7.27	&	215.96$\pm$47.37	&	0.82$\pm$0.11	&	270/238	&	6.62 	
&	2.80 	&	14.00$\pm$4.99	&	198.33$\pm$30.37	
&	0.91$\pm$0.07	&	262/239	\\
bn090717034	&	13.18 	&	15.8$\pm$4.29	&	186.17$\pm$4.73	&	1.07$\pm$0.01	&	326/237	&	32.08 	&	
14.27 	&	5.97$\pm$5.85	&	91.25$\pm$16.78	&	
0.71$\pm$0.14	&	241/238	\\
bn091030828	&	6.91 	&	8.29$\pm$0.36	&	110.8$\pm$13.43	&	0.12$\pm$0.13	&	245/238	&	10.67 	&	
20.93 	&	29.02$\pm$17.10	&	949.52$\pm$152.99	&	
1.01$\pm$0.04	&	266/238	\\
bn091120191	&	39.68 	&	18.00$\pm$3.61	&	181.36$\pm$20.12	&	1.16$\pm$0.05	&	232/239	&	9.17 	
&	5.89 	&	4.19$\pm$0.32	&	79.58$\pm$15.96	&	
1.00$\pm$0.14	&	254/239	\\
bn091123298	&	18.24 	&	4.93$\pm$1.46	&	154.59$\pm$21.22	&	0.57$\pm$0.10	&	234/240	&	
143.47 	&	20.35 	&	4.84$\pm$0.33	&	256.32$\pm$76.94	
&	1.38$\pm$0.09	&	222/240	\\
bn091208410	&	3.68 	&	4.01$\pm$0.19	&	101.4$\pm$30.94	&	1.13$\pm$0.16	&	275/237	&	5.42 	&	
3.58 	&	15.45$\pm$2.39	&	158.02$\pm$27.19	&	
1.12$\pm$0.08	&	259/237	\\
bn100116897	&	3.90 	&	2.69$\pm$0.58	&	138.45$\pm$52.16	&	0.66$\pm$0.23	&	274/240	&	80.00 	
&	19.33 	&	73.97$\pm$12.05	&	1233.3$\pm$139.74	
&	1.04$\pm$0.02	&	202/240	\\
bn100224112	&	33.86 	&	6.27$\pm$3.98	&	237.05$\pm$81.76	&	1.18$\pm$0.12	&	237/239	&	33.20 	
&	6.21 	&	1.78$\pm$0.91	&	132.07$\pm$92.9	&	
1.39$\pm$0.30	&	270/239	\\
bn100322045	&	12.13 	&	17.8$\pm$5.11	&	146.93$\pm$10.28	&	0.82$\pm$0.04	&	252/239	&	6.97 	
&	20.26 	&	57.31$\pm$14.2	&	665.03$\pm$67.29	
&	0.86$\pm$0.04	&	258/239	\\
bn100517072	&	3.71 	&	7.92$\pm$4.36	&	136.15$\pm$32.58	&	1.37$\pm$0.10	&	282/239	&	23.79 	
&	19.26 	&	2.03$\pm$0.84	&	281.59$\pm$191.22	
&	1.43$\pm$0.17	&	192/239	\\
bn100615083	&	19.58 	&	6.35$\pm$3.84	&	152.35$\pm$28.03	&	1.20$\pm$0.07	&	237/239	&	9.22 	
&	11.65 	&	5.65$\pm$3.27	&	124.76$\pm$42.16	
&	1.31$\pm$0.14	&	242/239	\\
bn100619015	&	10.11 	&	1.19$\pm$1.06	&	153.11$\pm$49.85	&	1.13$\pm$0.16	&	220/239	&	67.87 	
&	21.44 	&	4.29$\pm$0.30	&	146.35$\pm$51.86	
&	1.36$\pm$0.15	&	221/239	\\
bn100709602	&	11.33 	&	21.46$\pm$0.08	&	242.93$\pm$82.94	&	0.69$\pm$0.19	&	193/238	&	44.37 	
&	15.17 	&	2.04$\pm$0.81	&	198.64$\pm$80.66	
&	1.20$\pm$0.16	&	232/238	\\
bn100719989	&	6.78 	&	113.37$\pm$6.33	&	305.56$\pm$12.37	&	0.65$\pm$0.02	&	274/238	&	12.53 	
&	3.71 	&	24.43$\pm$11.34	&	420.3$\pm$100.79	
&	1.04$\pm$0.09	&	288/238	\\
bn101023951	&	16.96 	&	4.66$\pm$0.78	&	236.16$\pm$84.45	&	1.35$\pm$0.14	&	232/240	&	34.91 	
&	34.69 	&	60.15$\pm$5.79	&	276.52$\pm$20.67	
&	1.18$\pm$0.03	&	233/240	\\
bn101224578	&	8.96 	&	2.42$\pm$0.49	&	54.87$\pm$12.06	&	0.33$\pm$0.22	&	256/240	&	24.77 	&	
7.17 	&	1.88$\pm$0.39	&	74.1$\pm$32.99	&	
1.63$\pm$0.22	&	234/240	\\
bn101231067	&	8.58 	&	9.63$\pm$4.85	&	175.51$\pm$15.94	&	0.52$\pm$0.06	&	251/238	&	9.95 	
&	5.70 	&	10.04$\pm$2.95	&	65.13$\pm$6.12	&	
0.46$\pm$0.09	&	252/238	\\
bn110709463	&	4.86 	&	4.76$\pm$3.01	&	70.65$\pm$12.69	&	0.98$\pm$0.13	&	264/241	&	11.47 	&	
4.70 	&	9.61$\pm$3.34	&	126.51$\pm$19.88	&	
1.01$\pm$0.09	&	316/241	\\
bn110717319	&	22.91 	&	19.41$\pm$5.26	&	339.82$\pm$24.4	&	0.88$\pm$0.03	&	228/237	&	18.10 	&	
6.40 	&	7.03$\pm$0.50	&	486.16$\pm$120.68	&	
1.06$\pm$0.08	&	268/235	\\
bn110729142	&	44.54 	&	39.59$\pm$20.34	&	661.62$\pm$221.67	&	1.06$\pm$0.10	&	211/237	&	
116.05 	&	30.66 	&	5.54$\pm$1.10	&	177.87$\pm$30.13	
&	0.68$\pm$0.11	&	247/237	\\
bn110824009	&	6.98 	&	27.93$\pm$8.87	&	946.6$\pm$176.81	&	0.89$\pm$0.05	&	249/239	&	9.20 	
&	2.24 	&	2.83$\pm$0.47	&	832.71$\pm$148.8	
&	1.41$\pm$0.02	&	198/239	\\
bn110825102	&	12.29 	&	69.55$\pm$3.45	&	268.49$\pm$17.1	&	1.00$\pm$0.03	&	224/238	&	52.11 	&	
7.62 	&	2.81$\pm$0.70	&	100.18$\pm$33.97	&	
1.20$\pm$0.19	&	226/238	\\
bn110903009	&	7.23 	&	13.41$\pm$4.52	&	34.62$\pm$3.16	&	1.26$\pm$0.09	&	257/237	&	13.17 	&	
6.98 	&	10.75$\pm$4.30	&	305.11$\pm$82.31	&	
1.45$\pm$0.07	&	205/239	\\
bn110903111	&	21.89 	&	6.18$\pm$0.66	&	162.03$\pm$20.39	&	0.34$\pm$0.10	&	252/239	&	
167.84 	&	21.95 	&	5.73$\pm$0.24	&	352.62$\pm$58.03	
&	0.78$\pm$0.07	&	252/239	\\
bn110904124	&	2.56 	&	3.48$\pm$0.88	&	259.77$\pm$104.46	&	0.60$\pm$0.24	&	268/237	&	36.10 	
&	24.06 	&	3.79$\pm$0.34	&	412.63$\pm$153.23	
&	1.22$\pm$0.11	&	212/237	\\
bn110921912	&	8.58 	&	104.68$\pm$6.74	&	472.57$\pm$29.36	&	0.82$\pm$0.02	&	259/237	&	8.22 	
&	1.79 	&	32.79$\pm$6.28	&	1159.32$\pm$238	&	
1.07$\pm$0.04	&	232/237	\\
bn110926107	&	18.62 	&	1.74$\pm$0.49	&	207.54$\pm$98.64	&	1.30$\pm$0.15	&	215/238	&	39.24 	
&	17.09 	&	4.52$\pm$0.38	&	115.31$\pm$19.91	
&	0.96$\pm$0.10	&	181/238	\\
bn111024722	&	16.83 	&	8.17$\pm$4.28	&	139.27$\pm$18.48	&	0.97$\pm$0.07	&	220/239	&	32.53 	
&	19.78 	&	1.48$\pm$0.34	&	48.27$\pm$11.3	&	
0.78$\pm$0.22	&	205/239	\\
bn111228657	&	25.28 	&	12.04$\pm$2.16	&	410.98$\pm$234.65	&	1.94$\pm$0.08	&	203/237	&	31.07 	
&	15.10 	&	5.32$\pm$3.34	&	187.16$\pm$101.4	
&	2.19$\pm$0.10	&	228/236	\\
bn120412920	&	2.75 	&	2.37$\pm$0.52	&	39.09$\pm$1.15	&	0.15$\pm$0.04	&	273/239	&	68.27 	&	
21.57 	&	2.80$\pm$0.27	&	150.66$\pm$41.78	&	
1.22$\pm$0.11	&	214/239	\\
bn120530121	&	6.53 	&	3.02$\pm$1.66	&	126.29$\pm$31.11	&	0.30$\pm$0.19	&	219/240	&	43.98 	
&	13.44 	&	4.43$\pm$0.28	&	113.21$\pm$28.93	
&	1.09$\pm$0.13	&	211/240	\\
bn120605453	&	1.50 	&	11.19$\pm$4.20	&	270.23$\pm$64.97	&	1.00$\pm$0.10	&	268/237	&	12.06 	
&	5.79 	&	0.98$\pm$0.63	&	51.88$\pm$25.11	&	
1.30$\pm$0.35	&	211/237	\\
bn120618919	&	6.43 	&	2.23$\pm$0.36	&	108.23$\pm$33.41	&	0.90$\pm$0.20	&	212/238	&	7.55 	
&	3.78 	&	1.97$\pm$0.49	&	89.84$\pm$36.37	&	
0.44$\pm$0.34	&	263/238	\\
bn120711115	&	4.86 	&	5.73$\pm$2.13	&	285.49$\pm$18.56	&	0.50$\pm$0.04	&	238/241	&	58.98 	
&	48.96 	&	166.02$\pm$20.08	&	
1361.46$\pm$65.98	&	0.98$\pm$0.01	&	263/239	\\
bn120716712	&	6.21 	&	0.97$\pm$0.56	&	126.61$\pm$43.96	&	0.91$\pm$0.21	&	246/238	&	
170.83 	&	35.97 	&	4.14$\pm$0.30	&	153.19$\pm$21.49	
&	1.03$\pm$0.07	&	247/238	\\
bn121029350	&	0.45 	&	0.17$\pm$0.24	&	105.35$\pm$7.12	&	0.33$\pm$0.07	&	270/239	&	10.80 	&	
3.95 	&	8.58$\pm$0.25	&	119.23$\pm$11.33	&	
0.39$\pm$0.09	&	273/237	\\
bn121031949	&	9.41 	&	5.90$\pm$0.43	&	257.03$\pm$74.37	&	0.67$\pm$0.16	&	237/239	&	
182.40 	&	12.35 	&	1.45$\pm$0.45	&	209.17$\pm$82.21	
&	1.06$\pm$0.16	&	230/239	\\
bn121113544	&	16.90 	&	9.58$\pm$8.66	&	277.2$\pm$50.58	&	0.77$\pm$0.08	&	228/238	&	24.45 	&	
31.94 	&	4.84$\pm$0.36	&	120.75$\pm$36.2	&	
0.88$\pm$0.17	&	230/238	\\
bn121118576	&	12.35 	&	14.52$\pm$3.95	&	228.95$\pm$56.84	&	1.10$\pm$0.08	&	237/238	&	8.89 	
&	13.57 	&	1.06$\pm$0.48	&	102.95$\pm$45.91	
&	1.36$\pm$0.20	&	233/238	\\
bn121122870	&	27.01 	&	3.37$\pm$0.12	&	242.77$\pm$59.87	&	1.22$\pm$0.08	&	238/237	&	92.88 	
&	7.74 	&	2.82$\pm$0.33	&	99.11$\pm$21.35	&	
0.68$\pm$0.15	&	241/237	\\
bn121225417	&	27.20 	&	40.50$\pm$6.18	&	458.56$\pm$40.23	&	0.95$\pm$0.03	&	224/239	&	19.09 	
&	28.86 	&	24.14$\pm$4.19	&	287.27$\pm$30.52	
&	1.22$\pm$0.03	&	197/239	\\
bn130106995	&	13.50 	&	5.61$\pm$0.84	&	190.28$\pm$50.68	&	1.25$\pm$0.11	&	257/239	&	21.57 	
&	35.01 	&	6.80$\pm$6.32	&	133.63$\pm$28.99	
&	1.51$\pm$0.09	&	187/239	\\
bn130121835	&	24.26 	&	18.86$\pm$5.74	&	164.9$\pm$13.73	&	0.68$\pm$0.05	&	227/238	&	130.90 	&	
32.32 	&	1.78$\pm$0.74	&	184.51$\pm$91.3	&	
1.24$\pm$0.21	&	216/238	\\
bn130219775	&	1.79 	&	9.05$\pm$5.30	&	236.69$\pm$64.47	&	1.00$\pm$0.13	&	267/238	&	70.58 	
&	26.05 	&	23.90$\pm$4.88	&	638.77$\pm$126.38	
&	1.13$\pm$0.06	&	238/238	\\
bn130320560	&	16.70 	&	1.25$\pm$0.19	&	212.73$\pm$198.78	&	1.61$\pm$0.22	&	201/238	&	
141.34 	&	53.44 	&	6.47$\pm$6.47	&	281.2$\pm$111.44	
&	1.27$\pm$0.10	&	188/238	\\
bn130522510	&	2.03 	&	4.80$\pm$0.15	&	222.38$\pm$144.44	&	1.59$\pm$0.19	&	290/237	&	10.30 	
&	7.46 	&	3.43$\pm$3.39	&	70.36$\pm$16.63	&	
1.11$\pm$0.16	&	270/238	\\
bn130530719	&	11.52 	&	1.98$\pm$0.55	&	100.81$\pm$19.88	&	0.83$\pm$0.14	&	249/237	&	38.54 	
&	9.98 	&	1.72$\pm$0.55	&	347.79$\pm$439.28	
&	1.99$\pm$0.19	&	249/237	\\
bn130609902	&	30.08 	&	34.17$\pm$6.55	&	408.22$\pm$29.61	&	0.70$\pm$0.03	&	235/239	&	
144.64 	&	36.22 	&	1.49$\pm$0.69	&	308.3$\pm$244.42	
&	1.64$\pm$0.16	&	179/239	\\
bn130720582	&	80.26 	&	7.58$\pm$3.33	&	80.31$\pm$13.99	&	0.70$\pm$0.13	&	227/237	&	37.54 	&	
136.70 	&	19.38$\pm$2.34	&	98.87$\pm$11.25	&	
1.21$\pm$0.06	&	249/237	\\
bn130815660	&	3.07 	&	1.71$\pm$0.41	&	344.26$\pm$204.99	&	1.42$\pm$0.15	&	273/238	&	27.89 	
&	11.97 	&	22.30$\pm$2.33	&	93.52$\pm$7.4	&	
0.97$\pm$0.05	&	247/239	\\
bn130821674	&	57.22 	&	55.74$\pm$4.47	&	603.53$\pm$86.2	&	1.07$\pm$0.05	&	221/238	&	28.62 	&	
8.38 	&	8.57$\pm$3.81	&	165.24$\pm$58.81	&	
1.15$\pm$0.16	&	243/238	\\
bn131108024	&	8.83 	&	4.75$\pm$4.54	&	129.26$\pm$30.61	&	0.97$\pm$0.13	&	255/237	&	
223.36 	&	29.57 	&	4.61$\pm$1.06	&	200.48$\pm$60.42	
&	0.13$\pm$0.26	&	238/237	\\
bn140108721	&	11.58 	&	8.33$\pm$5.30	&	173.59$\pm$38.53	&	1.21$\pm$0.09	&	243/241	&	67.94 	
&	14.53 	&	12.72$\pm$5.87	&	479.54$\pm$110.13	
&	1.19$\pm$0.06	&	215/241	\\
bn140110814	&	9.22 	&	5.17$\pm$0.60	&	110.61$\pm$31	&	0.88$\pm$0.19	&	246/238	&	32.86 	&	
11.65 	&	1.26$\pm$1.08	&	65.18$\pm$37.39	&	
1.34$\pm$0.37	&	259/238	\\
bn140304849	&	8.26 	&	1.59$\pm$1.21	&	119.19$\pm$44.87	&	0.67$\pm$0.27	&	238/239	&	
187.97 	&	10.50 	&	3.41$\pm$0.51	&	78.02$\pm$19.3	&	
0.51$\pm$0.21	&	240/239	\\
bn140323433	&	64.38 	&	13.81$\pm$3.87	&	152.1$\pm$14.72	&	0.88$\pm$0.05	&	204/241	&	20.37 	&	
45.95 	&	7.14$\pm$3.18	&	98.86$\pm$16.19	&	
1.11$\pm$0.09	&	221/241	\\
bn140406120	&	44.29 	&	1.76$\pm$0.59	&	141.73$\pm$62.91	&	1.23$\pm$0.20	&	204/239	&	47.82 	
&	5.89 	&	1.80$\pm$0.43	&	121.19$\pm$55.38	
&	1.30$\pm$0.21	&	236/239	\\
bn140512814	&	6.27 	&	15.12$\pm$12.08	&	578.01$\pm$172.97	&	0.84$\pm$0.11	&	274/238	&	94.24 	
&	44.99 	&	12.01$\pm$8.56	&	992.34$\pm$347.95	
&	1.16$\pm$0.07	&	204/240	\\
bn140810782	&	53.06 	&	44.06$\pm$5.02	&	341.54$\pm$34.04	&	0.82$\pm$0.05	&	215/239	&	24.64 	
&	21.06 	&	16.9$\pm$4.83	&	175.65$\pm$36.69	
&	1.18$\pm$0.09	&	243/239	\\
bn140818229	&	9.41 	&	1.72$\pm$0.42	&	82.97$\pm$28.76	&	0.82$\pm$0.26	&	256/239	&	64.06 	&	
44.10 	&	17.18$\pm$7.97	&	204.42$\pm$16.78	&	
0.84$\pm$0.05	&	257/239	\\
bn140824606	&	11.46 	&	2.87$\pm$2.88	&	306.67$\pm$162.11	&	0.94$\pm$0.19	&	234/238	&	62.85 	
&	39.30 	&	9.84$\pm$6.60	&	183.89$\pm$34.16	
&	0.90$\pm$0.09	&	221/238	\\
bn140827763	&	11.07 	&	13.97$\pm$4.49	&	218.68$\pm$20.28	&	0.77$\pm$0.05	&	240/239	&	11.44 	
&	3.58 	&	3.06$\pm$0.22	&	105.86$\pm$5.37	&	
1.29$\pm$0.03	&	240/239	\\
bn150118409	&	38.34 	&	122.34$\pm$16.28	&	684.01$\pm$40.04	&	0.89$\pm$0.02	&	242/238	&	
6.11 	&	6.72 	&	109.32$\pm$9.23	&	
660.14$\pm$58.01	&	0.85$\pm$0.04	&	240/238	\\
bn150126868	&	13.25 	&	2.22$\pm$0.81	&	231.91$\pm$61.61	&	0.84$\pm$0.13	&	210/239	&	40.42 	
&	45.76 	&	13.6$\pm$4.85	&	296.76$\pm$53.17	
&	1.23$\pm$0.05	&	207/237	\\
bn150220598	&	38.91 	&	21.87$\pm$4.85	&	323.42$\pm$78.37	&	1.08$\pm$0.09	&	210/237	&	
103.42 	&	20.74 	&	4.27$\pm$0.49	&	249.88$\pm$87.1	&	
1.02$\pm$0.13	&	203/237	\\
bn150228981	&	8.06 	&	3.67$\pm$0.88	&	98.05$\pm$11.24	&	0.05$\pm$0.13	&	282/239	&	11.87 	&	
11.46 	&	4.84$\pm$2.61	&	140.23$\pm$25.12	&	
1.30$\pm$0.08	&	226/239	\\
bn150323395	&	11.97 	&	7.13$\pm$5.01	&	93.73$\pm$13.21	&	0.89$\pm$0.10	&	242/238	&	10.48 	&	
36.61 	&	4.08$\pm$0.46	&	131.27$\pm$23.76	&	
1.19$\pm$0.10	&	177/238	\\
bn150330828	&	11.46 	&	18.03$\pm$10.41	&	205.78$\pm$19.87	&	0.40$\pm$0.07	&	243/241	&	
104.98 	&	42.69 	&	87.1$\pm$5.10	&	313.04$\pm$21.55	
&	0.95$\pm$0.03	&	262/238	\\
bn150426594	&	4.90 	&	1.68$\pm$1.07	&	80.11$\pm$29.07	&	0.85$\pm$0.28	&	276/239	&	8.70 	&	
4.77 	&	24.11$\pm$5.28	&	193.56$\pm$26.66	&	
1.03$\pm$0.08	&	266/237	\\
bn150430015	&	35.33 	&	14.67$\pm$5.30	&	355.55$\pm$63.01	&	0.94$\pm$0.06	&	222/239	&	81.09 	
&	4.86 	&	2.04$\pm$0.55	&	227.07$\pm$152.04	
&	1.28$\pm$0.22	&	243/239	\\
bn150507026	&	23.23 	&	12.29$\pm$7.21	&	182.31$\pm$34.52	&	0.67$\pm$0.13	&	228/238	&	34.27 	
&	3.07 	&	3.80$\pm$0.47	&	127.75$\pm$54.36	
&	0.46$\pm$0.32	&	268/238	\\
bn150523396	&	14.78 	&	6.43$\pm$0.65	&	298.87$\pm$30.86	&	0.31$\pm$0.08	&	206/241	&	12.74 	
&	9.02 	&	36.04$\pm$12.74	&	323.9$\pm$23.21	&	
0.47$\pm$0.05	&	218/241	\\
bn150619287	&	14.27 	&	4.87$\pm$4.23	&	628.67$\pm$204.26	&	1.44$\pm$0.06	&	225/237	&	28.11 	
&	15.30 	&	8.22$\pm$2.98	&	669.27$\pm$255.97	
&	1.58$\pm$0.06	&	217/237	\\
bn150724782	&	6.53 	&	7.48$\pm$1.04	&	517.58$\pm$218.46	&	0.89$\pm$0.16	&	270/232	&	9.90 	
&	22.40 	&	23.73$\pm$9.28	&	481.49$\pm$44.53	
&	0.64$\pm$0.05	&	228/237	\\
bn150729517	&	6.27 	&	2.44$\pm$0.37	&	262.9$\pm$129.25	&	1.13$\pm$0.16	&	237/240	&	16.11 	
&	14.85 	&	19.47$\pm$9.38	&	532.82$\pm$91.6	&	
0.99$\pm$0.05	&	248/240	\\
bn151027166	&	22.78 	&	5.85$\pm$3.09	&	136.37$\pm$38.58	&	1.16$\pm$0.13	&	254/240	&	73.63 	
&	25.47 	&	8.42$\pm$7.03	&	440.95$\pm$130.76	
&	1.27$\pm$0.07	&	201/240	\\
bn151030999	&	11.01 	&	6.22$\pm$0.19	&	239.89$\pm$44.57	&	0.60$\pm$0.11	&	238/238	&	74.46 	
&	49.60 	&	17.52$\pm$3.82	&	323.04$\pm$32.33	
&	1.06$\pm$0.04	&	224/238	\\
bn151107851	&	23.04 	&	35.16$\pm$9.91	&	211.92$\pm$19.04	&	0.46$\pm$0.07	&	244/237	&	72.35 	
&	4.80 	&	2.60$\pm$0.28	&	182.73$\pm$68.43	
&	0.85$\pm$0.20	&	251/237	\\
bn151227218	&	4.22 	&	13.98$\pm$3.81	&	259.63$\pm$45.26	&	1.02$\pm$0.07	&	237/240	&	16.43 	
&	24.51 	&	40.42$\pm$4.23	&	474.57$\pm$39.67	
&	1.26$\pm$0.02	&	287/238	\\
bn151231443	&	17.09 	&	100.15$\pm$12.87	&	296.34$\pm$22.6	&	1.04$\pm$0.04	&	276/236	&	43.65 	
&	13.57 	&	49.77$\pm$19.26	&	157.72$\pm$10.61	
&	0.74$\pm$0.06	&	238/236	\\
bn160215773	&	19.33 	&	4.68$\pm$0.55	&	410.44$\pm$173.17	&	1.00$\pm$0.14	&	212/238	&	88.56 	
&	32.70 	&	81.46$\pm$14.99	&	633.37$\pm$60.65	
&	0.87$\pm$0.03	&	248/238	\\
bn160225809	&	5.95 	&	4.63$\pm$0.26	&	74.11$\pm$17.21	&	0.31$\pm$0.25	&	260/237	&	37.97 	&	
22.72 	&	10.74$\pm$2.83	&	142.91$\pm$18.91	&	
0.94$\pm$0.08	&	251/237	\\
bn160303201	&	13.57 	&	9.74$\pm$0.59	&	312.27$\pm$120.22	&	1.34$\pm$0.11	&	231/238	&	11.64 	
&	13.38 	&	12.76$\pm$5.42	&	143.67$\pm$16.74	
&	1.02$\pm$0.06	&	249/238	\\
bn160325291	&	17.34 	&	17.98$\pm$5.39	&	190.66$\pm$16.89	&	0.65$\pm$0.05	&	238/238	&	27.49 	
&	3.71 	&	8.59$\pm$4.61	&	242.7$\pm$52.58	&	
0.87$\pm$0.10	&	239/238	\\
bn160802259	&	6.78 	&	119.40$\pm$4.56	&	258.19$\pm$11.62	&	0.60$\pm$0.03	&	295/241	&	8.70 	
&	4.03 	&	17.87$\pm$2.06	&	152.77$\pm$12.95	
&	0.93$\pm$0.04	&	284/239	\\
bn161220605	&	10.69 	&	10.64$\pm$4.25	&	150.26$\pm$18.5	&	0.79$\pm$0.07	&	245/236	&	18.55 	&	
3.90 	&	2.17$\pm$0.66	&	96.85$\pm$37.02	&	
0.82$\pm$0.26	&	177/236	\\
bn170115662	&	5.18 	&	1.40$\pm$0.25	&	89.42$\pm$29.56	&	0.68$\pm$0.26	&	283/237	&	80.38 	&	
15.68 	&	1.77$\pm$0.24	&	192.8$\pm$79.65	&	
1.23$\pm$0.16	&	236/237	\\
bn170130510	&	14.98 	&	9.21$\pm$5.53	&	206.56$\pm$73.27	&	1.17$\pm$0.14	&	222/237	&	44.06 	
&	46.34 	&	2.44$\pm$0.76	&	300.03$\pm$135.92	
&	1.20$\pm$0.15	&	198/238	\\
bn170208553	&	13.44 	&	6.38$\pm$0.05	&	545.8$\pm$170.1	&	1.06$\pm$0.10	&	214/241	&	24.00 	&	
11.39 	&	8.31$\pm$4.25	&	144.54$\pm$16.85	&	
0.56$\pm$0.09	&	228/241	\\
bn170209048	&	8.06 	&	8.73$\pm$6.02	&	153.71$\pm$27.37	&	0.88$\pm$0.10	&	243/237	&	20.14 	
&	11.07 	&	4.63$\pm$0.30	&	82.85$\pm$11.42	&	
0.77$\pm$0.10	&	226/237	\\
bn170228773	&	8.26 	&	2.33$\pm$0.46	&	300.66$\pm$114.79	&	1.17$\pm$0.13	&	275/239	&	10.79 	
&	6.59 	&	2.54$\pm$0.43	&	557.07$\pm$213.2	
&	1.27$\pm$0.09	&	261/239	\\
bn170317666	&	10.62 	&	2.95$\pm$2.57	&	225.14$\pm$47.72	&	0.49$\pm$0.16	&	257/238	&	
170.18 	&	19.58 	&	1.26$\pm$0.75	&	169.24$\pm$64.35	
&	1.18$\pm$0.17	&	201/238	\\
bn170409112	&	53.38 	&	52.87$\pm$1.76	&	796.69$\pm$42.21	&	0.74$\pm$0.02	&	314/237	&	16.48 	
&	25.79 	&	1.86$\pm$0.58	&	49.94$\pm$9.07	&	
0.39$\pm$0.20	&	224/241	\\
bn170423719	&	34.82 	&	8.59$\pm$2.81	&	230.34$\pm$28.89	&	1.36$\pm$0.04	&	251/239	&	13.02 	
&	14.40 	&	1.25$\pm$0.60	&	119.23$\pm$67.05	
&	1.33$\pm$0.26	&	232/239	\\
bn170510217	&	32.00 	&	25.39$\pm$6.45	&	412.85$\pm$37.63	&	0.98$\pm$0.03	&	235/237	&	71.42 	
&	32.00 	&	1.73$\pm$0.60	&	362.54$\pm$184.73	
&	1.30$\pm$0.15	&	197/237	\\
bn170514180	&	8.77 	&	7.19$\pm$5.33	&	517.55$\pm$212.53	&	1.30$\pm$0.09	&	240/238	&	69.50 	
&	33.86 	&	11.53$\pm$3.71	&	284.34$\pm$59.86	
&	1.57$\pm$0.05	&	192/238	\\

\enddata
\tablenotetext{a}{Duration of the first and second emission episodes.}

\tablenotetext{b}{Flux of the first and second emission episodes.}

\tablenotetext{c}{Peak energy of CPL model fits of the first and second emission episodes.}

\tablenotetext{d}{The low energy photon index of CPL model fits for the first and second emission
episodes.} \tablenotetext{e}{The quiescent times that are calculated from the end of the first episode
to the beginning of the second episode.}

\tablenotetext{1}{The parameters of the Band function fits to the second episode of GRB 150330A are
$\alpha=(0.95\pm 0.03)$, $\beta=(2.3\pm0.07)$, $E_{\rm p}=(313\pm22)$, and $\chi^2=1.1$; and for the
first episode of GRB 170409A the parameters are $\alpha=(0.74\pm0.02)$, $\beta=(2.67\pm0.11)$, $E_{\rm
p}=(797\pm42)$, and $\chi^2=1.32$.}

\end{deluxetable}

\clearpage

\begin{figure}
\begin{tabular}{lllll}
\multirow{2}{*}{ \includegraphics[angle=0,scale=0.45]{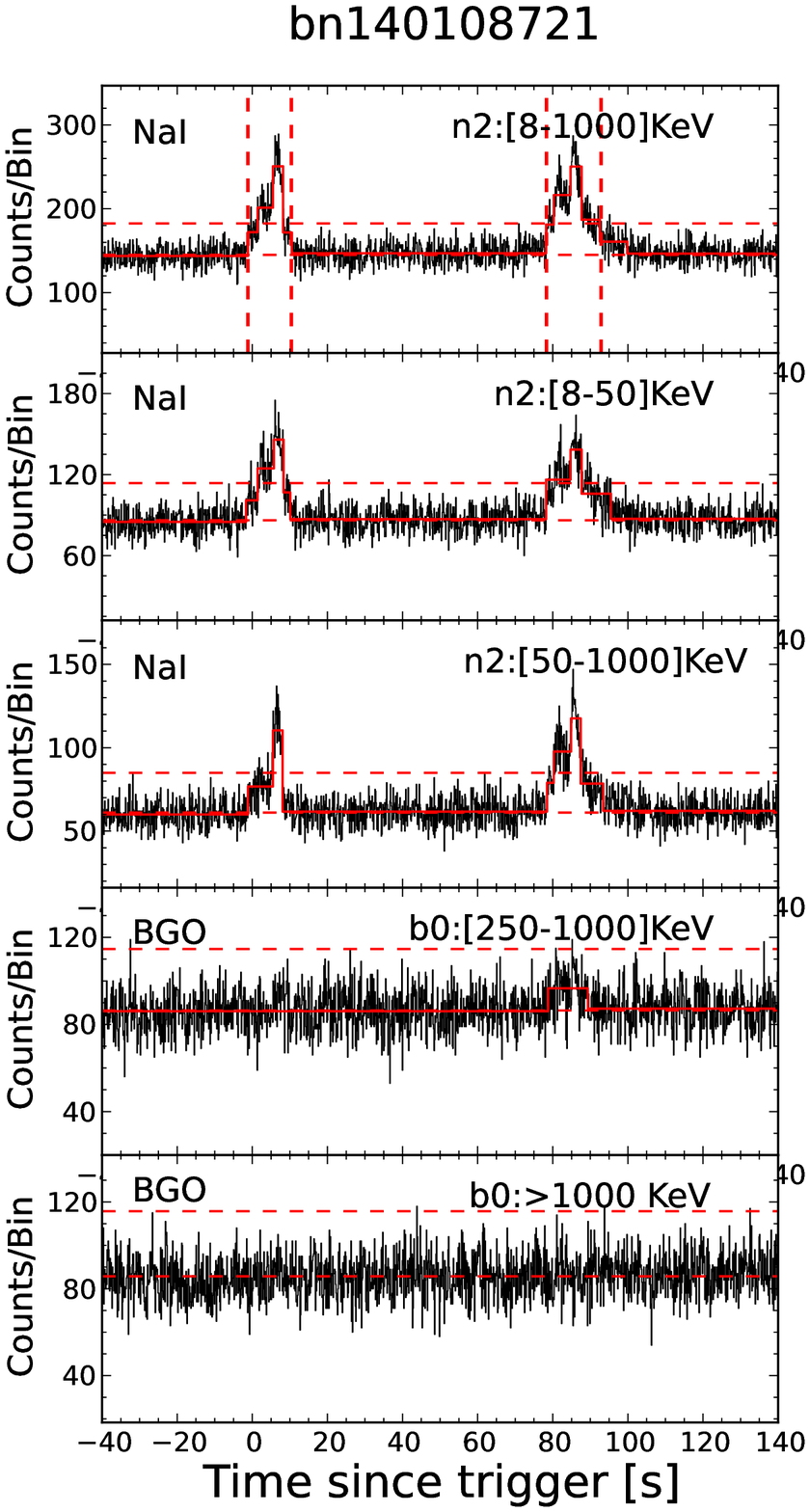}} \\
&\includegraphics[angle=270,scale=0.28]{bn140108721_1.eps} \\
&\includegraphics[angle=270,scale=0.28]{bn140108721_2.eps} \\
\end{tabular}
\center{Fig. \ref{LC2example}---Continued.}
\end{figure}


\end{document}